\documentclass[reprint, aps,pra,twocolumn,superscriptaddress]{revtex4-1}
\usepackage{braket}
\usepackage{amssymb}
\usepackage{amsmath}
\usepackage{epsfig}
\usepackage{color}
\usepackage{graphics, graphicx}
\usepackage{bbold}
\usepackage{psfrag}
\usepackage{mathcomp}
\usepackage{subfigure}
\usepackage{verbatim}
\usepackage{physics}
\usepackage[colorlinks, citecolor=blue]{hyperref}
\usepackage[normalem]{ulem}
\usepackage{bm}
\def\cp#1{\mathbf{#1}}

\begin{document}

\title{Effective scatterings and universal clusters of heteronuclear ultracold mixtures in quasi-low dimensions}
\author{Tingting Shi}
\affiliation{Beijing National Laboratory for Condensed Matter Physics, Institute of Physics, Chinese Academy of Sciences, Beijing, 100190, China}
\affiliation{Beijing Academy of Quantum Information Sciences, Beijing 100093, China}
\author{Xiaoling Cui}
\email{xlcui@iphy.ac.cn}
\affiliation{Beijing National Laboratory for Condensed Matter Physics, Institute of Physics, Chinese Academy of Sciences, Beijing, 100190, China}
\date{\today}

\begin{abstract} 
We study the effective s-wave scattering of two heteronuclear atoms harmonically confined in quasi-low dimensions, where the atoms have unequal masses and are subject to different confinement frequencies. The resulting effective scattering parameters in low dimensions, including scattering length and  effective range, are derived as functions of  three-dimensional scattering parameters and  confinement strengths. Using realistic Li-K and Li-Cr mixtures as examples, we further compute the binding energies of universal $(1+N)$ clusters in quasi-low dimensions using the effective scattering parameters. Our findings suggest a promising pathway for practically observing  universal clusters and their associated many-body phases in low-dimensional ultracold heteronuclear systems.

\end{abstract}
\maketitle

\section{Introduction}

Ultracold atoms provide an ideal platform for the quantum simulation of strongly interacting low-dimensional systems, which exhibit distinct quantum phenomena from their three-dimensional (3D) counterparts \cite{review1, review2}. In this context, external confinement potentials generated by magnetic and optical traps have emerged as an efficient tool for tuning both dimensionality and scattering resonances. A well-known example is the confinement-induced resonance (CIR) in quasi-one dimension (q1D) \cite{Olshanii1, Olshanii2}, which demonstrates that a new s-wave scattering resonance and bound state can emerge by tuning the s-wave scattering length relative to the transverse confinement length. The predicted CIR and induced bound state have been successfully observed in ultracold gases confined in q1D \cite{ETH, sTG_Innsbruck}, facilitating experimental explorations of strongly repulsive 1D atomic gases in the hard-core and super-Tonks-Girardeau regimes \cite{sTG_Innsbruck, Jochim1, Jochim2, Lev, Weiss1, Weiss2}. In recent years, CIR has been extended to various systems for engineering strong interactions, including quasi-two dimension (q2D) \cite{Petrov1,Petrov2}, optical lattices \cite{Fedichev, Zhou, Buchler}, mixed dimensions or systems with asymmetric/anharmonic potentials \cite{mixed_D, mixed_D_expt, NJP, Schmelcher1, Schmelcher2, Schmelcher3,Cui_FBS, split_CIR_expt, Peng, Zhang, split_CIR_theory}, high partial-wave scattering \cite{CIR_p_expe, CIR_p_1,CIR_p_2,CIR_p_3,CIR_p_4},  multiple spin channels \cite{Cui2, Pietro, ZhangWei, Liu, ZhangWei2, Blume_soc}, etc.

Recently, highly tunable interactions have also been realized in heteronuclear mixtures with unequal masses \cite{K_Li1, K_Li2, K_Li3, K_Li4, K_Li5, Cr_Li, Cr_Li2,Cr_Li3, Cr_Li4,Dy_K1, Dy_K2, Dy_K3, Dy_K4}. A unique property of such systems is their ability to support universal $(1+N)$ clusters, consisting of one light atom and $N \ (\ge 2)$ heavy identical fermions, which are insensitive to short-range interaction details and are therefore collisionally stable. In 3D, the critical heavy-light mass ratio required to support these universal bound states is quite high \cite{KM, Blume, Petrov}, but it can be significantly reduced in 2D and 1D \cite{Pricoupenko, Parish, Cui, Liu_PRR, KM_1D, Mehta_1D,Petrov_1D}, making them accessible in systems such as $^{6}$Li-$^{40}$K \cite{K_Li1, K_Li2, K_Li3, K_Li4, K_Li5}, $^{6}$Li-$^{53}$Cr \cite{Cr_Li, Cr_Li2,Cr_Li3, Cr_Li4}, and $^{40}$K-$^{161}$Dy \cite{Dy_K1, Dy_K2, Dy_K3, Dy_K4}. These clusters also suggest the emergence of intriguing new many-body phases dominated by high-order correlations \cite{Parish3, Parish4, Naidon, mass_polaron, QSF,Petrov_N1,Petrov_N2}, such as the trimer liquid \cite{Naidon} and quartet superfluid \cite{QSF}. However, to practically explore such universal clusters and their induced many-body phases, one must first understand the basic two-body scattering of mass-imbalanced mixtures in low dimensions. In particular, due to the heteronuclear nature, heavy and light atoms generally experience different trapping frequencies, i.e., species-selective confinement. Such asymmetric trapping leads to strong coupling between the center-of-mass (COM) and relative motions of the two atoms \cite{Cui_FBS, NJP, Schmelcher1, Schmelcher2, Schmelcher3, mixed_D}, thereby introducing considerable complexity into the study of CIR in these systems. Another practical issue to consider is the presence of large effective range in the two-body scattering of these heteronuclear systems, as they are generally associated with very narrow Feshbach resonances. For example, for commonly used Feshbach resonances in $^{6}$Li-$^{40}$K \cite{review_LiK_KDy} and $^{6}$Li-$^{53}$Cr \cite{review_LiCr} systems, the effective range can be on the order of $10^3 a_0$ (with $a_0$ the Bohr radius), comparable to or even larger than the scattering length in unitary regime. How such a considerable effective range modifies low-dimensional scattering under species-selective confinement is a practically important question to address.

In this work, we study the effective scattering of two heteronuclear atoms in both q1D and q2D, where different atoms are associated with different harmonic trapping frequencies. By diagonalizing a large matrix expanded over different COM levels, we extract low-dimensional scattering parameters, including scattering length and effective range, as functions of 3D scattering parameters and confinement strength. These effective parameters govern the interaction of heteronuclear atoms in reduced low dimensions. Based on these scattering parameters, we further compute the binding energies of universal $(1+N)$ clusters for realistic $^{6}$Li-$^{40}$K and $^{6}$Li-$^{53}$Cr mixtures in q1D and q2D. 
Our results suggest a promising future for the practical detection of universal clusters and their associated many-body phases in low-dimensional ultracold heteronuclear systems.

The remainder of this paper is organized as follows. In Section~\ref{model}, we present the model and formalism for the effective scattering of two heteronuclear atoms in reduced low dimensions, where the COM and relative motions of the two atoms are coupled due to their different trapping frequencies. In Section~\ref{results}, we show our numerical results for low-dimensional scattering parameters in realistic Li-K and Li-Cr systems in both q1D and q2D, based on which we further calculate the binding energies of universal clusters and identify optimal conditions for their detection. Finally, we summarize this work in Section~\ref{summary}.

\section{Model and Formalism} \label{model}

We first write down the Hamiltonian of two heteronuclear atoms under harmonic confinement: ($\hbar=1$)
\begin{equation}
H=-\frac{\nabla^2_{\cp r_h}}{2m_h}-\frac{\nabla^2_{\cp r_l}}{2m_l} +V_h({\cp r_h})+V_l({\cp r_l})+g\delta({\cp r_h}-{\cp r_l}), 
\end{equation}
where ${\cp r_h}$ (${\cp r_l}$) is the coordinate of heavy (light) atom, with mass $m_h$ ($m_l$); $V_h$ and $V_l$ are their confinement potentials that generate quasi-low geometries; $g$ is the bare s-wave coupling in 3D, which can be related to the scattering length $a_s$ via $1/g=\mu/(2\pi a_s)-1/V \sum_{\mathbf{Q}}2\mu/\mathbf{Q}^2$, with $V$ the 3D volume and $\mu\equiv m_hm_l/(m_h+m_l)$ the reduced mass, where $\mathbf{Q}$ denotes the 3D momentum. In the presence of a finite effective range $R^*$, $a_s$ becomes energy-dependent~\cite{Chin}:
\begin{equation}
a_s^{-1}(E) = a_s^{-1} + R^* (2\mu E), \label{as_E}
\end{equation} 
with $E$ the two-body collision energy  in COM frame.

Next, we derive the effective two-body scatterings of heteronuclear atoms in q2D and q1D, which are generated by longitudinal and transverse harmonic confinements, respectively. Here, we focus on the case where different atoms experience different trapping frequencies and their COM and relative motions are strongly coupled with each other. Throughout the paper, we consider the heteronuclear atoms near a single s-wave resonance with highly tunable $a_s$, while neglecting their possible p-wave interaction that may lead to non-trivial scattering features due to the confinement-induced s- and p-wave hybridization\cite{Schmelcher1, Schmelcher2}. Moreover, we will also neglect the confinement-induced three-body interaction in effective low dimensions\cite{prl2008, 1D_3_body_1, 1D_3_body_2, 1D_3_body_3}, given that such a three-body interaction is strongly suppressed in the current $(1+2)$ system due to the Pauli-blocking effect of identical fermions.

\subsection{Q2D}  \label{q2d}

To create q2D geometry, we consider a longitudinal harmonic trap:
\begin{equation}
V_{\sigma}({\cp r}_{\sigma})=\frac{1}{2}m_{\sigma}\omega_{\sigma}^2 z_{\sigma}^2;\ \ \ \ \ (\sigma=h,l)
\end{equation}
where $z_{\sigma}$ is the longitudinal component of ${\cp r}_{\sigma}$,  and $\omega_{\sigma}$ is the trapping frequency for $\sigma$-atom. Since the transverse COM motion can be decoupled from the problem, we just consider three relevant coordinates of two-body system: two longitudinal coordinates ($z_h,\ z_l$) and a relative coordinate in $xy$ plane (${\bm \rho}\equiv {\bm \rho}_h-{\bm \rho}_l$). Accordingly, the eigen-state of non-interacting Hamiltonian can be denoted as $|n_h, n_l; {\cp k}\rangle$, with eigen-energy $E_{n_h, n_l; {\cp k}}=\epsilon^h_{n_h}+\epsilon^l_{n_l}+{\cp k}^2/(2\mu)$ and $\epsilon^{\sigma}_{n_{\sigma}}=(n_{\sigma}+1/2)\omega_{\sigma}$.

For low-energy scattering at long distances, the heavy and light atoms can move freely only in the transverse ($xy$) direction, as their longitudinal motions along $z$ are frozen by the harmonic confinement. This corresponds to an effective 2D system. To study the effective 2D scattering, we consider the incident state
\begin{equation}
|\Psi^{(0)}\rangle=|n_h=n_l=0; {\cp q}\rangle,
\end{equation}
where $n_h=n_l=0$ denote the harmonic ground state along $z$, and ${\cp q}$ is the relative wave vector in $xy$ plane. The incident energy is then $E_{0,0; {\cp q}}=(\omega_h+\omega_l)/2+q^2/(2\mu)$ (here  $q\equiv |{\cp q}|$), and we assume $q^2/(2\mu)\ll \omega_h, \ \omega_l$ such that no longitudinal excitation is allowed for effective 2D scattering. Based on the Lippmann-Schwinger equation, the scattered state follows
\begin{equation}
|\Psi\rangle=|\Psi^{(0)}\rangle+G_0T|\Psi^{(0)}\rangle, \label{LS}
\end{equation}
where $G_0=(E-H_0+i\delta)^{-1}$ is the Green's function for non-interacting Hamiltonian $H_0$, and $T$ is the scattering matrix. At long distance $|\bm{\rho}|\rightarrow\infty$, the wavefunction (\ref{LS}) in real space can be reduced  to  													
\begin{equation}
\langle z_h, z_l, \bm{\rho}|\Psi\rangle\rightarrow \phi_0^h(z_h)\phi_0^l(z_l) \left( J_0(q\rho) -\frac{i\mu}{2} t_q H_0(q\rho)\right),
\end{equation}																	with
\begin{equation}
t_q= S\langle \Psi^{(0)}| T |\Psi^{(0)}\rangle. \label{t1}
\end{equation}
Here, $J_0(x)$ and $H_0(x)$ are the Bessel function of the first kind and the Hankel function of the first kind, respectively. 
Furthermore, $t_q$ is the T-matrix element that determines the effective 2D scattering:
\begin{equation}
t_q\equiv\frac{2\pi}{\mu}  \frac{1}{-\ln(q^2a_{\rm 2D}^2)+i\pi+R_{\rm 2D}q^2}, \label{t2}
\end{equation}	
where $a_{\rm 2D}$ and $R_{\rm 2D}$ are respectively the 2D scattering length and effective range. To obtain these effective parameters, the key question is to solve $t_q$ from (\ref{t1}). 

To solve $t_q$, we introduce an auxiliary‌ state $|f\rangle\equiv T|\Psi^{(0)}\rangle$ and  express
\begin{equation}
t_q= S\langle \Psi^{(0)} | f \rangle. \label{t_q}
\end{equation}																	Given $T=U+UG_0U+UG_0UG_0U+...=(1-UG_0)^{-1}U$, where $U=g\delta({\cp r})$ (with ${\cp r}\equiv {\cp r}_h-{\cp r}_l$) is the bare interaction, we have
\begin{equation}
|f\rangle=\left( \frac{1}{g}-\delta({\cp r}) G_0\right)^{-1} \delta({\cp r}) |\Psi^{(0)}\rangle
\end{equation}																	or equivalently
\begin{equation}
\left( \frac{1}{g}-\delta({\cp r}) G_0\right)|f\rangle=\delta({\cp r}) |\Psi^{(0)}\rangle.
\end{equation}																	Expanding $G_0$ in terms of the eigen-states of $H_0$, we get
\begin{equation}
\left( \frac{1}{g}-\delta({\cp r}) \sum_{n_h, n_l; {\cp k}} \frac{ | n_h, n_l; {\cp k} \rangle \langle n_h, n_l; {\cp k} |}{E-E_{n_h,n_l;{\cp k}}+i\delta} \right)|f\rangle=\delta({\cp r}) |\Psi^{(0)}\rangle. \label{eq}
\end{equation}	
For the summation in the bracket, the contribution from $n_h=n_l=0$ gives $\sum_{\cp k}\frac{2\mu}{q^2-{\cp k}^2+i\delta}|\Psi^{(0)}\rangle \langle \Psi^{(0)}|=S\frac{\mu}{2\pi}(-\ln\Lambda^2+\ln q^2 -i\pi) |\Psi^{(0)}\rangle \langle \Psi^{(0)}|$, with $\Lambda$ the cutoff of transverse momentum amplitude.	Note that in above we have used $\delta({\cp r}) | 0,0; {\cp k} \rangle =\delta({\cp r}) | \Psi^{(0)}\rangle$ for any momentum ${\cp k}$, which is a unique property for contact interaction. In this way, we can simplify (\ref{eq}) as
\begin{equation}
\left( \frac{1}{g}-\delta({\cp r}) A  \right)|f\rangle=\delta({\cp r}) \left(1+ t_q \frac{\mu}{2\pi} (\ln \tilde{q}^2-i\pi) \right) |\Psi^{(0)}\rangle, \label{eq2}
\end{equation}	
with
\begin{equation}
A=\sum_{{n_hn_l}\neq{00}; {\cp k}} \frac{ | n_h, n_l; {\cp k} \rangle \langle n_h, n_l; {\cp k} |}{E-E_{n_h,n_l;{\cp k}}} -  \frac{\mu}{2\pi} S\ln\tilde{\Lambda}^2 |\Psi^{(0)}\rangle \langle \Psi^{(0)}|.
\end{equation}	
Here we have rescaled momenta as $\tilde{\Lambda}=\Lambda l_0$ and $\tilde{q}=ql_0$, with $l_0$ the length unit.
From (\ref{eq2}), we can further get 
\begin{equation}
\frac{1}{t_q}=\frac{\mu}{2\pi} (i\pi -\ln \tilde{q}^2) +\langle \Psi^{(0)}| \left[ \frac{1}{g}-\delta({\cp r}) A  \right]^{-1}  \delta({\cp r}) S |\Psi^{(0)}\rangle^{-1}. \label{tq2}
\end{equation}
Comparing (\ref{tq2}) with (\ref{t2}), we then have
\begin{eqnarray}
&&\frac{\mu}{2\pi} \left(-\ln \tilde{a}_{\rm 2D}^2 +R_{\rm 2D}q^2 \right)  \nonumber\\
&=&  \langle \Psi^{(0)}| \left[ \frac{1}{g}-\delta({\cp r}) A  \right]^{-1}  \delta({\cp r}) S |\Psi^{(0)}\rangle ^{-1}, \label{final_eq}
\end{eqnarray}
from which one can extract both $a_{2D}$ and $R_{2D}$. Next we show how to solve the right-hand side of above equation.

Considering the structure of the rhs of (\ref{final_eq}), we employ a molecular basis $\{|N\rangle \equiv |N\rangle_Z |\bm{\rho}=0\rangle\}$ to expand the matrix 
\begin{equation}
{ \cal M}= \frac{1}{g}-\delta({\cp r}) A. \label{M}
\end{equation}
In principle, one can choose $\{|N\rangle_Z\}$ as any complete eigenstates of COM motion along $z$. Here, to improve the efficiency of numerical calculations, we  choose $|N\rangle_Z$ as the eigenstate of following Hamiltonian 
\begin{equation}
H_M=-\frac{1}{2M} \frac{\partial^2}{\partial Z^2} + \frac{1}{2} M \omega_M^2 Z^2,
\label{HM_2d}
\end{equation}
with 
\begin{equation}
M=m_h+m_l,\ \ \ \ \omega_M=\sqrt{\frac{m_h\omega_h^2+m_l\omega_l^2}{M}}. \label{M}
\end{equation}
Obviously, $H_M$ is the COM Hamiltonian of our system when heavy and light atoms are forced to stay at the same site, i.e., representing an artificial molecule\cite{Cui_FBS}. 
In this way, the matrix element of ${\cal M}$ can be expressed as
\begin{eqnarray} 
	{\cal M}_{NN'} &=& \frac{1}{g} \delta_{NN'}-  \frac{1}{S} \sum_{\{n_hn_l\}\neq\{00\}; {\cp k}} \frac{  F_{N; n_h,n_l} F^*_{N'; n_h,n_l}}{E-E_{n_h,n_l;{\cp k}}} \nonumber\\
	&&+  \frac{\mu}{2\pi} \ln\tilde{\Lambda}^2 F_{N; 0,0} F^*_{N'; 0,0} \label{M_element}
\end{eqnarray} 
with 
\begin{equation}
F_{N; n_h,n_l}=\bra{N}\ket{n_h,n_l;{\bf k}}
=\int dZ \Phi^*_N(Z) \phi^h_{n_h}(Z)\phi^l_{n_l}(Z).
\end{equation} 
Here $ \Phi_N$ is the eigenstate of $H_M$  with eigen-energy $\epsilon_N=(N+1/2)\omega_M$, and $\phi^{\sigma}_n$ ($\sigma=h,l$) is the eigenstate of single-particle Hamiltonian for $\sigma$-atom. To incorporate the finite range effect, one needs to impose an energy-dependence on $a_s$ (see Eq.\ref{as_E}),  which can be done by replacing $\frac{1}{g}\delta_{NN'}$ in (\ref{M_element}) by
\begin{equation}
\frac{1}{g}\delta_{NN'}\rightarrow \left( \frac{\mu}{2\pi} \Big[\frac{1}{a_s} + R^*(2\mu E_N)\Big] -\frac{1}{V}\sum_{\cp Q} \frac{2\mu}{{\cp Q}^2}  \right) \delta_{NN'} \label{range}
\end{equation}
with $E_N=q^2/(2\mu)+\omega_h/2+\omega_l/2-\epsilon_N$.  It can be analytically shown that the 3D ultraviolet divergence in $1/g$ term can be exactly canceled by the divergence from last two terms in (\ref{M_element}), see appendix \ref{appendix_q2D}. 

The diagonalization of ${\cal M}$ matrix gives 
\begin{eqnarray} 
	{\cal M}|\alpha\rangle_j = \left(\frac{\mu}{2\pi a_s}+\lambda_j\right) |\alpha\rangle_j,
\end{eqnarray} 
with $|\alpha\rangle_j$ the $j$-th eigenstate and $\frac{\mu}{2\pi a_s}+\lambda_j$ the according eigenvalue. Explicitly, we can express $|\alpha\rangle_j=\sum_{N}c_{jN}|N\rangle$, with $\{c_{jN}\}$ the expansion coefficients. Finally, the rhs of (\ref{final_eq}) can be evaluated as 
\begin{eqnarray} 
	&&\left[\sum_{j} \frac{| \sum_N c_{jN}F_{N;00} |^2}{\frac{\mu}{2\pi a_s}+\lambda_j}\right]^{-1}. \label{rhs}
\end{eqnarray} 
The expansion of (\ref{rhs}) at small $q$ gives the zeroth and quadratic terms as in (\ref{final_eq}), which respectively determine the effective 2D parameters $a_{\rm 2D}$ and $R_{\rm 2D}$.
In the numerical calculation, we truncate the molecular COM basis
$\{|N\rangle\}$ at $N\leq N_{\max}=10$. We have explicitly checked the convergence with respect to $N_{\max}$ and find that the effective scattering parameters are  well converged for the basis size adopted here.

Our formalism can also produce the bound state solution, which corresponds to the divergence of T-matrix, i.e., $t_q=\infty$. Specifically, the two-body binding energy $E_2=-\kappa^2/(2\mu)$ can be obtained by replacing $q^2\rightarrow -\kappa^2$ in (\ref{tq2}) and sending $1/t_q$ to zero. On the other hand, one can utilize the effective 2D parameters to solve the bound state. In this case, we recall Eq.(\ref{t2}) and the divergence of $t_q$ leads to 
\begin{equation}
\ln(\kappa^2a_{\rm 2D}^2)+R_{\rm 2D}\kappa^2=0. \label{E2_2d}
\end{equation}
The comparison of two $E_b$ (or $\kappa$) in above methods can be used as a criterion for judging the validity of effective 2D description based on $a_{\rm 2D}$ and $R_{\rm 2D}$.

Once  the effective 2D parameters are obtained, one can solve the universal clusters from Skorniakov–Ter-Martirosian (STM) equations in 2D\cite{Pricoupenko, Parish, Cui, Liu_PRR}. 
Here we consider $(1+N)$ clusters consisting of one light atom and $N$
identical heavy fermions, experiencing only two-body $s$-wave interaction (between heavy-light atoms) but not any p-wave interaction (between heavy-heavy or heavy-light) or three-body interaction (between heavy-heavy-light or heavy-light-light). 
Explicitly, we define the momentum-space wavefunction of ($1+N$) clusters in the dimer-fermion frame as $f_{{\cp k}_2,...{\cp k}_N}$, which describes a heavy-light dimer with momentum $-\sum_{i=2}^N{\cp k}_i$ and $N-1$ heavy fermions with momenta $ {\cp k}_2,...,{\cp k}_N$. In the presence of a finite range $R_{\rm 2D}$, the resulting STM equation reads\cite{Liu_PRR}
\begin{eqnarray} 
	&&f_{\bf k_2\cdots k_N}\left[
	\frac{\mu^2S}{\pi} R_{\rm 2D} {\cal A}_{{\cp k}_2...{\cp k}_N}  
	- \sum_{\bf k} \frac{2\mu}{k^2+a_{\rm 2D}^{-2}} 
	  +  \sum_{\bf k_1} \frac{1}{E_{\bf k_1 k_2\cdots k_N}}
	\right]\notag\\
	&&=\sum_{\bf k_1} \frac{\sum_{i=2}^N f_{\bf k_2\cdots k_{i-1}k_ik_{i+1}\cdots k_N}\delta_{\bf kk_i}}{E_{\bf k_1 k_2 \cdots k_N}},  \label{q2d_cluster}
\end{eqnarray} 
where $\varepsilon_{\bf k}^{\sigma}=\frac{{\cp k}^2}{2m_{\sigma}}$ ($\sigma=h,l$), $\varepsilon_{\bf k}^d=\frac{{\cp k}^2}{2(m_h+m_l)}$, $E_{\bf k_1 k_2 \cdots k_N}=-E_{1+N} + \sum_{i=1}^N\varepsilon_{\bf k_i}^h +\varepsilon_{\bf k_1+k_2\cdots + k_N}^l$, ${\cal A}_{{\cp k}_2...{\cp k}_N}=E_{1+N} - \sum_{i=2}^N\varepsilon_{\bf k_i}^h - \varepsilon_{\bf k_2\cdots + k_N}^d$. By solving above equation, one can obtain the binding energy $E_{1+N}$ of $(1+N)$ cluster in 2D.

\subsection{Q1D}  \label{q1d}

In this section, we study the effective scattering of heavy and light atoms confined in transverse  harmonic potentials:
\begin{equation}
V_{\sigma}({\cp r}_{\sigma})=\frac{1}{2}m_{\sigma}\omega_{\sigma}^2 \bm{\rho}_{\sigma}^2;\ \ \ \ \ (\sigma=h,l)
\end{equation}
with $\bm{\rho}_{\sigma}=(x_{\sigma}, y_{\sigma})$ the transverse component of ${\cp r}_{\sigma}$. Separating out the longitudinal COM motion, we consider three relevant coordinates of this two-body system: two transverse coordinates ($\bm{\rho}_h,\ \bm{\rho}_l$) and a relative coordinate along longitudinal direction ($z\equiv z_h-z_l$). Accordingly, the eigen-state of non-interacting Hamiltonian can be denoted as $|{\cp n}_h, {\cp n}_l; k\rangle$, with eigen-energy $E_{{\cp n}_h, {\cp n}_l; k}=\epsilon^h_{{\cp n}_h}+\epsilon^l_{{\cp n}_l}+k^2/(2\mu)$.

At long distances and low energies, the transverse motions of the heavy and light atoms are frozen in the harmonic ground state, allowing only longitudinal motion. This effective 1D system leads us to consider the incident state
\begin{equation}
|\Psi^{(0)}\rangle=|{\cp n}_h={\cp n}_l={\cp 0}; q\rangle,
\end{equation}
where ${\cp n}_h={\cp n}_l={\cp 0}$ denote the ground state in transverse direction, and $q$ is the relative wave vector along z. The incident energy is then $E_{0,0; {\cp q}}=\omega_h+\omega_l+q^2/(2\mu)$, and we assume $q^2/(2\mu)\ll \omega_h, \ \omega_l$. Following the Lippmann-Schwinger equation as in (\ref{LS}), the scattered state in real space at long distance $|z|\rightarrow\infty$ can be reduced to  									
\begin{equation}
\langle \bm{\rho}_h, \bm{\rho}_l, z|\Psi\rangle\rightarrow \phi_0^h(\bm{\rho}_h)\phi_0^l(\bm{\rho}_l) \left(\cos(qz) +\frac{\mu}{iq} t_q e^{iq|z|} \right),
\end{equation}																	with
\begin{equation}
t_q= L\langle \Psi^{(0)}| T |\Psi^{(0)}\rangle. \label{t1_1D}
\end{equation}
Here $t_q$ is the T-matrix element that determines the effective 1D scattering:
\begin{equation}
t_q\equiv-\frac{1}{\mu}  \frac{1}{a_{1D}-i/q-R_{1D}q^2}, \label{t2_1D}
\end{equation}	
where $a_{1D}$ and $R_{1D}$ are respectively the 1D scattering length and effective range. 


Similar to the strategy of solving $t_q$ in quasi-2D case, we introduce an auxiliary‌ state $|f\rangle\equiv T|\Psi^{(0)}\rangle$ and  express
\begin{equation}
t_q= L\langle \Psi^{(0)} | f \rangle. \label{t_q_1D}
\end{equation}
Then we have
\begin{equation}
\left( \frac{1}{g}-\delta({\cp r}) \sum_{{\cp n}_h, {\cp n}_l; k} \frac{ | {\cp n}_h, {\cp n}_l; k \rangle \langle {\cp n}_h, {\cp n}_l; k |}{E-E_{{\cp n}_h, {\cp n}_l; k}+i\delta} \right)|f\rangle=\delta({\cp r}) |\Psi^{(0)}\rangle. \label{eq_1D}
\end{equation}	
For the summation in the bracket, the contribution from ${\cp n}_h={\cp n}_l=0$ gives $\sum_{k}\frac{2\mu}{q^2-k^2+i\delta}|\Psi^{(0)}\rangle \langle \Psi^{(0)}|=L\frac{\mu}{iq}|\Psi^{(0)}\rangle \langle \Psi^{(0)}|$. Then (\ref{eq_1D}) can be simplified as
\begin{equation}
\left( \frac{1}{g}-\delta({\cp r}) A  \right)|f\rangle=\delta({\cp r}) \left(1+ t_q \frac{\mu}{iq}  \right) |\Psi^{(0)}\rangle, \label{eq2_1D}
\end{equation}	
with
\begin{equation}
A=\sum_{{{\cp n}_h,{\cp n}_l}\neq{{\cp 0,\cp 0}}; k} \frac{ | {\cp n}_h, {\cp n}_l; k \rangle \langle {\cp n}_h, {\cp n}_l; k |}{E-E_{{\cp n}_h, {\cp n}_l; k}}.
\end{equation}	
We can see that, unlike in the q2D case, scattering along the free direction here does not produce an ultraviolet divergence. Consequently, $A$ is purely contributed from higher transverse modes but not the ground state ${\cp n}_h={\cp n}_l={\cp 0}$.

From Eq.(\ref{eq2_1D}), we can get $t_q$ as
\begin{equation}
\frac{1}{t_q}=-\frac{\mu}{iq} +\langle \Psi^{(0)}| \left[ \frac{1}{g}-\delta({\cp r}) A  \right]^{-1}  \delta({\cp r}) L |\Psi^{(0)}\rangle^{-1}. \label{tq2_1D}
\end{equation}
Comparing (\ref{tq2_1D}) with (\ref{t2_1D}), we have
\begin{equation}
-\mu \left(a_{\rm 1D}-R_{\rm 1D}q^2 \right) =\langle \Psi^{(0)}| \left[ \frac{1}{g}-\delta({\cp r}) A  \right]^{-1}  \delta({\cp r}) L |\Psi^{(0)}\rangle ^{-1}, \label{final_eq_1D}
\end{equation} 
from which one can extract both $a_{\rm 1D}$ and $R_{\rm 1D}$.

To expand the matrix ${ \cal M}= \frac{1}{g}-\delta({\cp r}) A$, we employ a molecular basis $|{\cp N}\rangle \equiv |{\cp N}\rangle_{\rho} |z=0\rangle$, where  $|{\cp N}\rangle_{\rho}$ is chosen as the eigen-state of transverse COM motion governed by the molecular Hamiltonian
\begin{equation}
H_M=-\frac{\nabla^2_{\bm{\rho}}}{2M} + \frac{1}{2} M\omega_M^2 \bm{\rho}^2,
\label{HM_1d}
\end{equation}
with $M$ and $\omega_M$ the same as presented in (\ref{M}). The effect of finite effective range can be included by modifying $1/g$ similarly as Eq.\ref{range}, where $E_{\cp N}$ is given by $E_{\cp N}=q^2/(2\mu)+\omega_h+\omega_l-\epsilon_{\cp N}$ and $\epsilon_{\cp N}=(2N_r+|M_a|+1)\omega_M$ is the eigen-energy of Hamiltonian $H_M$ with quantum number ${\cp N}=(N_r, M_a)$. 
A detailed evaluation of ${\cal M}$ matrix element is given in appendix \ref{appendix_q1D}. Due to the rotational invariance of our system, the angular momentum quantum number is conserved as $M_a=0$ for the ground state scattering. In our diagonalization of ${\cal M}$ matrix, we have truncated the radial quantum number as $N_r\leq 5$ and  confirmed the numerical convergence with this cutoff. We note that a similar q1D scattering problem has been studied previously \cite{NJP}; however, the finite-range effect was not discussed there, and moreover, the basis states chosen for the center-of-mass motion differ from ours.

After the diagonalization of ${\cal M}$ matrix, one can obtain the right-hand side of (\ref{final_eq_1D}). Its  expansion at small $q$ give the effective 1D parameters $a_{1D}$ and $R_{1D}$. 
Similar to q2D case, one can solve the two-body binding energy $E_2=-\kappa^2/(2\mu)$ by replacing $q^2\rightarrow -\kappa^2$ and sending $1/t_q$ to zero in (\ref{tq2_1D}). Alternatively, $E_b$ can also be solved from effective 1D parameters, i.e., by sending $t_q\rightarrow\infty$ in (\ref{t2_1D}), which gives   
\begin{equation}
a_{\rm 1D}-1/\kappa+R_{\rm 1D}\kappa^2=0. \label{E2_1d}
\end{equation} 

Based on  the effective 1D parameters, one can solve universal ($1+N$) clusters from the 1D STM equations\cite{Petrov_1D}. Similar to the 2D case, we can define the wavefunction of ($1+N$) cluster in dimer-fermion frame as $f_{{ k}_2,...{ k}_N}$, and finally arrive at the 1D STM equation that incorporates the finite range effect:
\begin{eqnarray} 
	&&f_{k_2\cdots k_N}\left[
	2\mu^2 L R_{\rm 1D}{\cal A}_{{k}_2...{k}_N}  -\mu L a_{\rm 1D}  +  \sum_{k_1} \frac{1}{E_{k_1 k_2\cdots k_N}}
	\right]\notag\\
	&&=\sum_{k_1} \frac{\sum_{i=2}^N f_{k_2\cdots k_{i-1}k_ik_{i+1}\cdots k_N}\delta_{kk_i}}{E_{k_1 k_2 \cdots k_N}},  \label{q1d_cluster}
\end{eqnarray} 
where $\varepsilon_{k}^{\sigma}=\frac{k^2}{2m_{\sigma}}$ ($\sigma=h,l$), $\varepsilon_{k}^d=\frac{k^2}{2(m_h+m_l)}$, $E_{k_1 k_2 \cdots k_N}=-E_{1+N} + \sum_{i=1}^N\varepsilon_{k_i}^h +\varepsilon_{k_1+k_2\cdots + k_N}^l$, ${\cal A}_{{k}_2...{k}_N}=E_{1+N} - \sum_{i=2}^N\varepsilon_{k_i}^h - \varepsilon_{k_2\cdots + k_N}^d$. The binding energy $E_{1+N}$ of this 1D system can be obtained from above equation.

\section{Results} \label{results}

In this section, we present our numerical results for effective low-dimensional scattering and universal $(1+N)$ binding of heteronuclear mixtures.  To be relevant to ongoing ultracold experiments, we consider two realistic mass-imbalanced fermionic mixtures. The first is a mixture of $^{6}$Li $|f=1/2, m_f=1/2\rangle $ and $^{53}$Cr $|f=9/2, m_f=-9/2\rangle$  near Feshbach resonance at $B_0=1414G$, which has mass ratio $m_h/m_l=8.8$ and 3D effective range $R^*\approx6000a_0$\cite{review_LiCr}. The second is a mixture of $^{6}$Li $|f=1/2, m_f=1/2\rangle $ and $^{40}$K $|f=9/2, m_f=-5/2\rangle$  near Feshbach resonance at $B_0=155G$,  with $m_h/m_l=6.7$ and $R^*\approx2400a_0$\cite{review_LiK_KDy}.  For both systems,    $R^*$ is sufficiently large that it must be explicitly taken into account in scattering and bound-state problems. Moreover, in these heteronuclear systems, the heavy and light atoms generally experience  different trapping frequencies, i.e., $\omega_h\neq \omega_l$. As a specific example, we take $\omega_h/\omega_l=m_l/m_h$, so that different atoms have the same confinement length $l_{\rm ho}=(m_h\omega_h)^{-1/2}=(m_l\omega_l)^{-1/2}$. In our numerics, we choose different values of $l_{\rm ho}$ ranging from $1500a_0$ to $700a_0$.

Fig.~\ref{fig_LiCr_q2d} shows the effective scattering and bound states of  Li-Cr mixture in q2D. In Fig.~\ref{fig_LiCr_q2d}(a), we observe multiple resonances of $a_{\rm 2D}$, which can be attributed to the involvement of multiple COM levels in the scattering process \cite{Cui_FBS, NJP, mixed_D}. 
In the weak-coupling regime with negative $a_s$, our results indicate that a smaller $l_{\text{ho}}$ (i.e., tighter confinement) leads to a smaller $a_{\rm 2D}$, corresponding to a deeper two-body bound state. In this regime, $R_{\rm 2D}$ is nearly a constant that depends on $R^{*}$ and $l_{\rm ho}$  but not on $a_s$, as shown in Fig.~\ref{fig_LiCr_q2d}(b). In Fig.~\ref{fig_LiCr_q2d}(c), the discrete data show the exact results of two-body binding energy $E_2$, which can be well fit by predictions from effective 2D model (Eq.~(\ref{E2_2d}), solid lines) over a wide range of negative $R^*/a_s$. This demonstrates the validity of effective 2D description in this regime.

Given the validity of effective 2D model, we further utilize the effective 2D parameters to compute the binding energies  of universal $(1+N)$ clusters (denoted by $E_{1+N}$) based on Eq.(\ref{q2d_cluster}), with $N=2,3$. To efficiently compare different clusters, we define the relative binding energy of $(1+N)$ cluster as \cite{Liu_PRR}
\begin{equation}
\Delta_{1+N,N}\equiv E_{1+N}-E_N.
\end{equation}
In Fig.~\ref{fig_LiCr_q2d}(c), we show $\Delta_{1+N,N}$ ($N=2$, $3$) as functions of $R^*/a_s$ at several values of $l_{\rm ho}$. Unlike the behavior of $E_2$, $\Delta_{1+N,N}$ varies non-monotonically with $R^*/a_s$, consistent with our previous findings for equal confinement frequencies~\cite{Liu_PRR}. In contrast, the ratio $|\Delta_{1+N,N}/E_2|$, shown in Fig.~\ref{fig_LiCr_q2d}(d), changes monotonically with $R^*/a_s$. In the weak-coupling limit ($R^*/a_s \rightarrow -\infty$), $|\Delta_{1+N,N}/E_2|$ saturates at its maximum value, which is determined by Eq.~(\ref{q2d_cluster}) with zero effective range ($R_{\rm 2D}/a_{\rm 2D}^2 \rightarrow 0$)~\cite{Liu_PRR} and therefore depends only on the mass ratio $m_h/m_l$. As $R^*/a_s$ increases, $|\Delta_{1+N,N}/E_2|$ decreases continuously from its maximum value to zero, i.e., $\Delta_{1+N,N} \rightarrow 0$, at which point the $(1+N)$ cluster energy merges into the continuum spectrum of an $N$-body cluster and a light atom.

To experimentally detect the $(1+N)$ cluster in realistic settings, it is necessary that both $|\Delta_{1+N,N}|$ and $|\Delta_{1+N,N}/E_2|$ be sufficiently large, so that the cluster’s spectrum can be clearly resolved from those of smaller clusters. Thus, an optimal condition is to employ strong confinement (i.e., small $l_{\rm ho}$) while choosing an intermediate $R^*/a_s$ where both quantities are visibly large. For example, in Fig.~\ref{fig_LiCr_q2d}(c,d), a favorable condition for detecting $(1+2)$ trimer in Li-Cr system is achieved at $l_{\rm ho}=700a_0$ within $R^*/a_s\in (-69.9, -61.8)$, where $\Delta_{32}/h$ ranges from -0.28 kHz to -0.92 kHz, and $|\Delta_{32}/E_2|$ lies between 27$\%$ and 10$\%$. In comparison, detecting the $(1+3)$ tetramer is more challenging due to its  smaller $|\Delta_{43}|$ and $|\Delta_{43}/E_2|$ relative to the trimer case. For instance, in the same interval of $R^*/a_s$ as above, we find $\Delta_{43}/h\in (-0.18, -0.50)$ kHz and $|\Delta_{43}/E_2|\in (17\%, 5\%)$.

We now turn to Fig.~\ref{fig_LiK_q2d} for Li-K system, which hosts a smaller $R^*$ than Li-Cr case. The behaviors of $a_{\rm 2D}$, $R_{\rm 2D}$, $E_2$, $\Delta_{1+N,N}$, and $|\Delta_{1+N,N}/E_2|$ in Fig.~\ref{fig_LiK_q2d}(a–d) are qualitatively similar to those shown in Fig.~\ref{fig_LiCr_q2d}(a–d). However, due to the smaller mass ratio in Li-K, the saturation value of $|\Delta_{1+N,N}/E_2|$ in the weak-coupling limit, as displayed in Fig.~\ref{fig_LiK_q2d}(d), is lower than that in Fig.~\ref{fig_LiCr_q2d}(d) for Li-Cr. Nevertheless, the deepest bindings of $\Delta_{1+N,N}$ in Fig.~\ref{fig_LiK_q2d}(c) are very close to those in Fig.~\ref{fig_LiCr_q2d}(c). For Li-K at $l_{\rm ho}=700a_0$, the favorable condition to observe q2D trimers and tetramers lies within $R^*/a_s\in(-15.3, -12.4)$, where $\Delta_{32}/h\in (-0.17, -0.80)$kHz,  $\Delta_{43}/h\in (-0.10, -0.42)$kHz, leading to $|\Delta_{32}/E_2|\in(16\%, 10\%)$, and $|\Delta_{43}/E_2|\in(9\%, 6\%)$.

\begin{widetext}

\begin{figure}[h]     
	\centering
	\includegraphics[height=10cm]{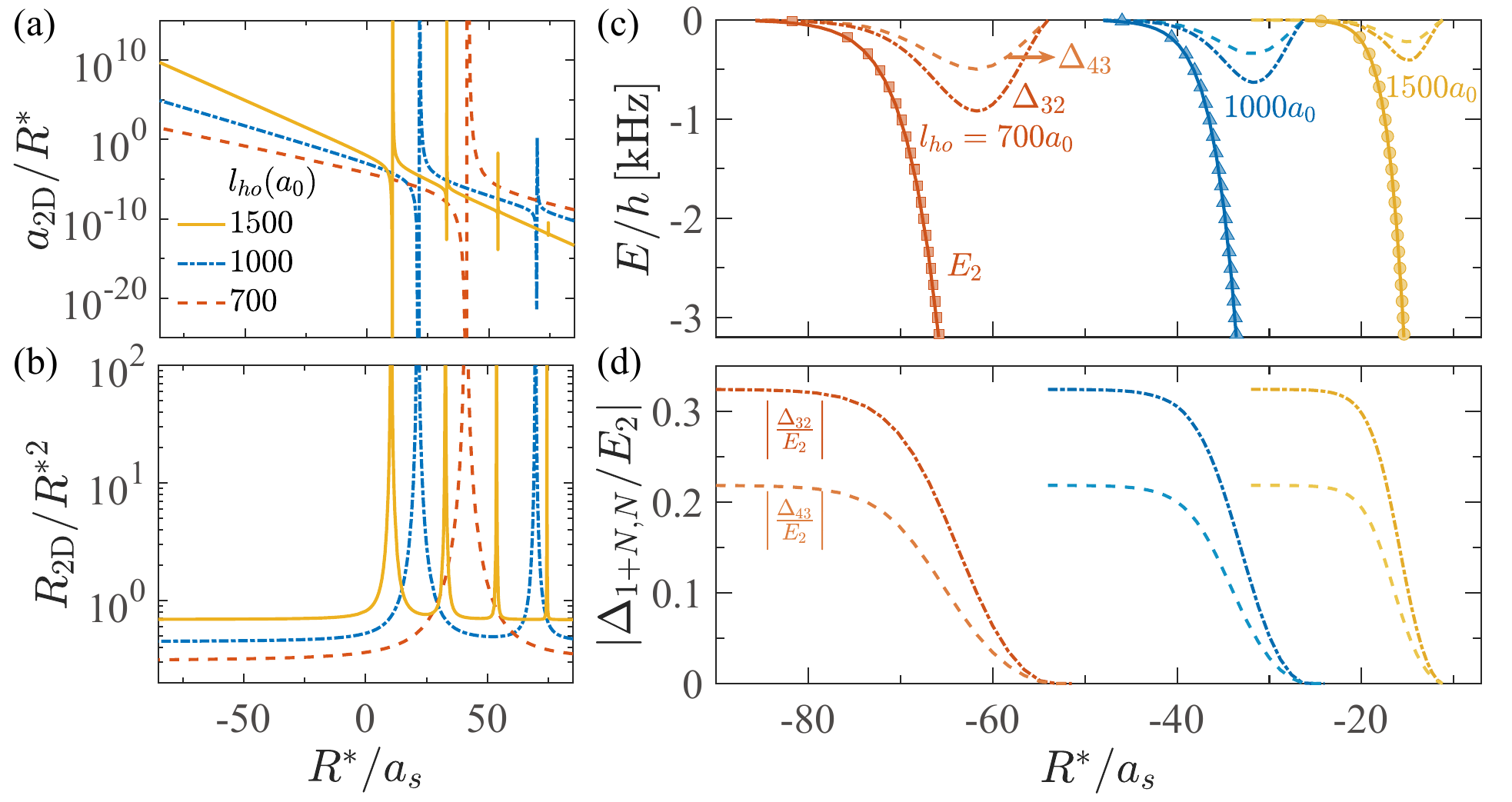}
	\caption{Quasi-2D effective scattering and bound states for $^6$Li-$^{53}$Cr mixture, under longitudinal harmonic confinement with equal confinement length $l_{\text{ho}} = (m_h \omega_h)^{-1/2} = (m_l \omega_l)^{-1/2}$. (a, b) The reduced 2D scattering length $a_{\rm 2D}$ and effective range $R_{\rm 2D}$  as functions of $R^*/a_s$. Here, $a_s$ and $R^* = 6000a_0$ are respectively the 3D scattering length and effective range.  (c) Two-body binding energy $E_2$ and relative binding energies $\Delta_{1+N,N} \equiv E_{1+N} - E_N$ of $(1+N)$ clusters (with $N=2,3$) as functions of $R^*/a_s$. For $E_2$, discrete data and solid line are respectively from exact solution and effective 2D model; For $\Delta_{32}$ and $\Delta_{43}$, they are all from effective 2D model. (d) Relative ratios $|\Delta_{1+N,N}/E_2|$ ($N=2,3$) as functions of $R^*/a_s$. In all  plots, three distinct confinement lengths are considered: $l_{\text{ho}} = 1500,\ 1000,\ 700\ (a_0)$.
	 }
	\label{fig_LiCr_q2d} 
	\end{figure}

\begin{figure}[h]     
	\centering
	\includegraphics[height=10cm]{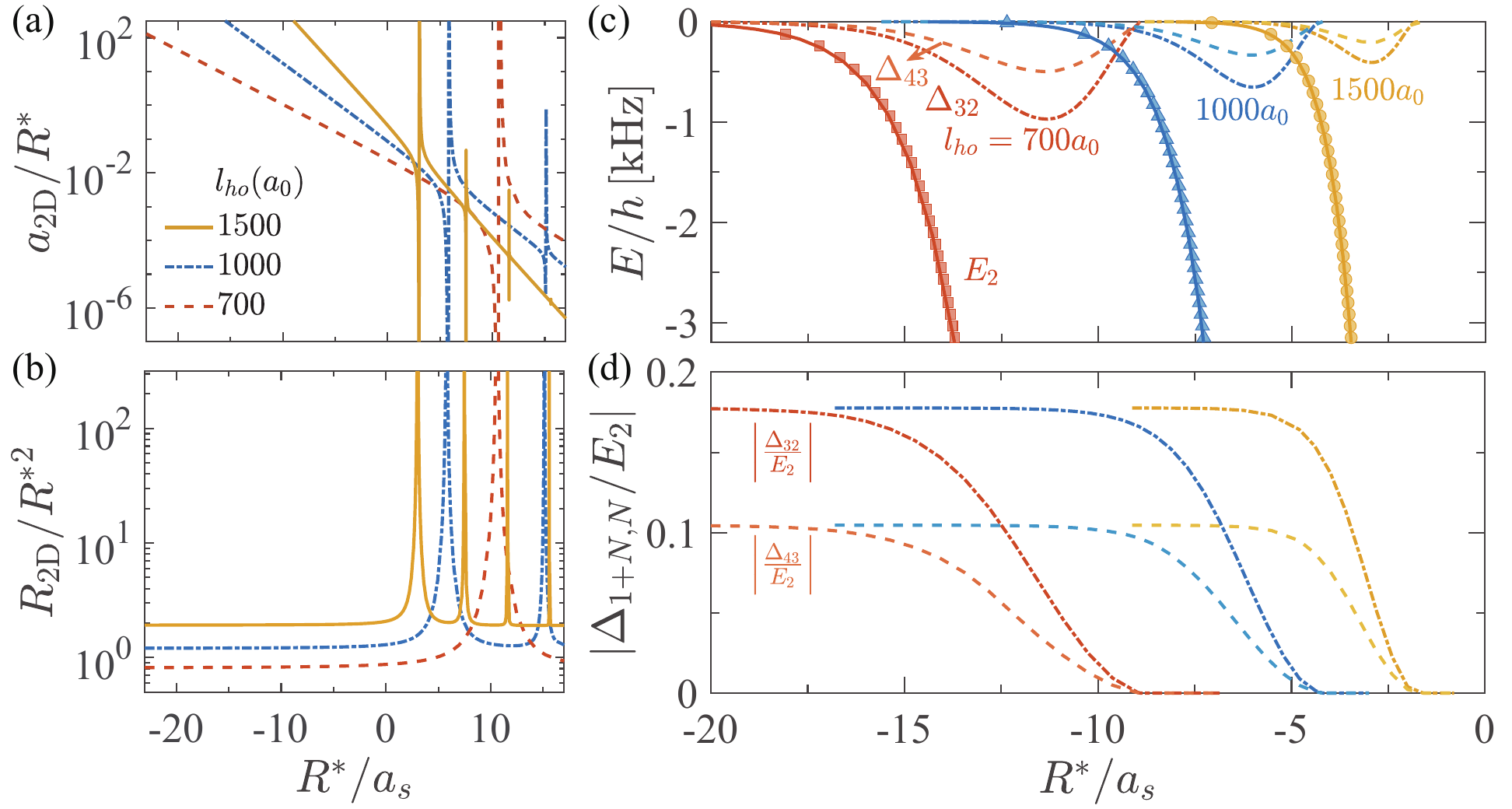}
	\caption{The same as Fig.\ref{fig_LiCr_q2d} except for $^6$Li-$^{40}$K mixture with 3D effective range $R^*=2400a_0$.  }
	\label{fig_LiK_q2d} 
\end{figure}



\begin{figure}[h]     
	\centering
	\includegraphics[height=10cm]{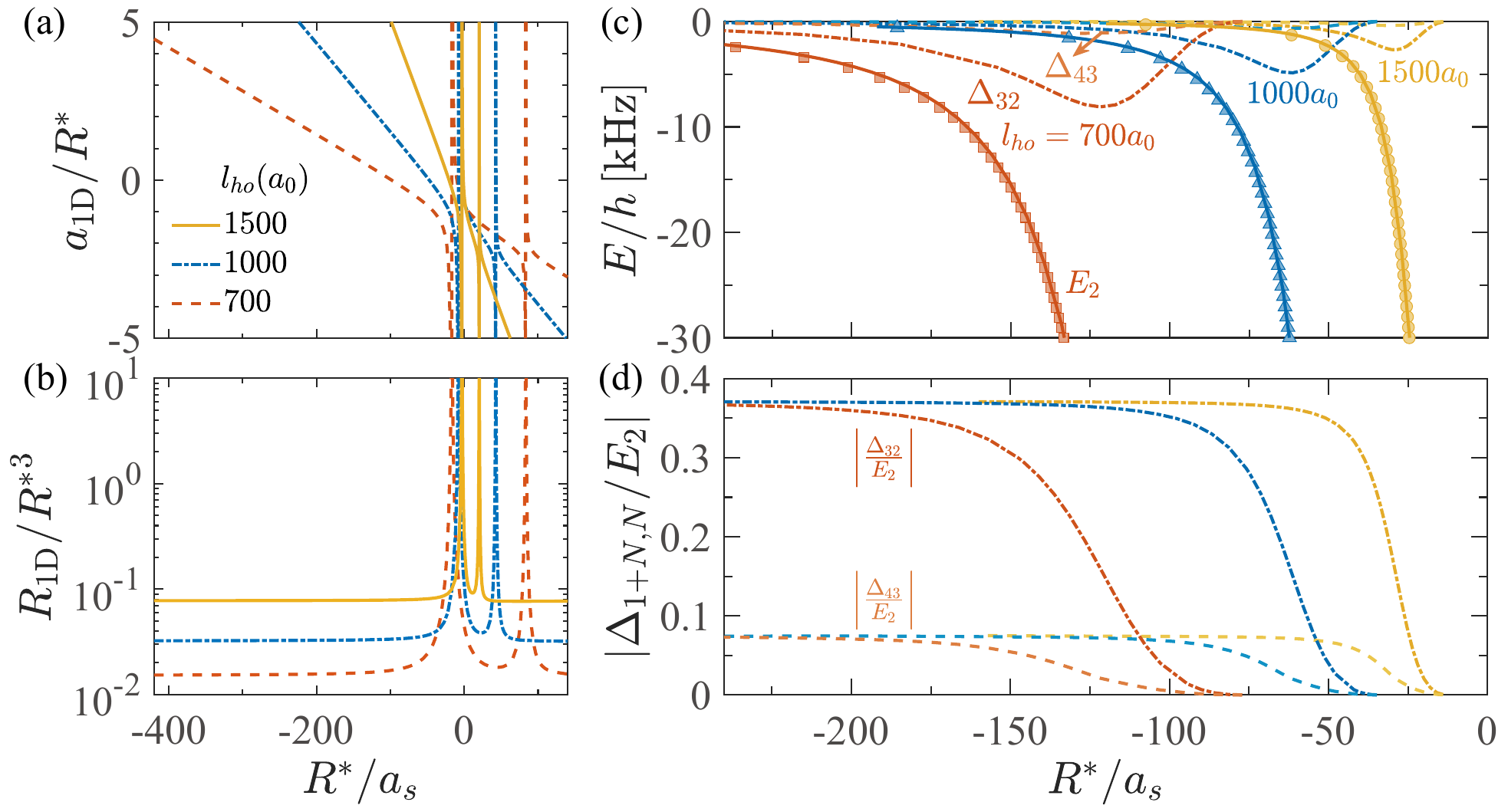}
	\caption{Quasi-1D effective scattering and bound states for $^6$Li-$^{53}$Cr mixture, under transverse harmonic confinement with equal confinement length $l_{\text{ho}} = (m_h \omega_h)^{-1/2} = (m_l \omega_l)^{-1/2}$. 	(a, b) The reduced 1D scattering length $a_{\rm 1D}$ and effective range $R_{\rm 1D}$ as functions of $R^*/a_s$.  (c) Two-body binding energy $E_2$ and relative binding energies $\Delta_{1+N,N} \equiv E_{1+N} - E_N$ of $(1+N)$ clusters (with $N=2,3$) as functions of $R^*/a_s$. For $E_2$, discrete data and solid line are respectively from exact solution and effective 1D model; For $\Delta_{32}$ and $\Delta_{43}$, they are all from effective 1D model. (d) Relative ratios $|\Delta_{1+N,N}/E_2|$ ($N=2,3$) as functions of $R^*/a_s$. In all  plots,  3D effective range $R^*=6000a_0$ and we take three different $l_{\text{ho}} =1500,\ 1000,\ 700 (a_0)$.  }
	\label{fig_LiCr_q1d} 
\end{figure}

\begin{figure}[h]     
	\centering
	\includegraphics[height=10cm]{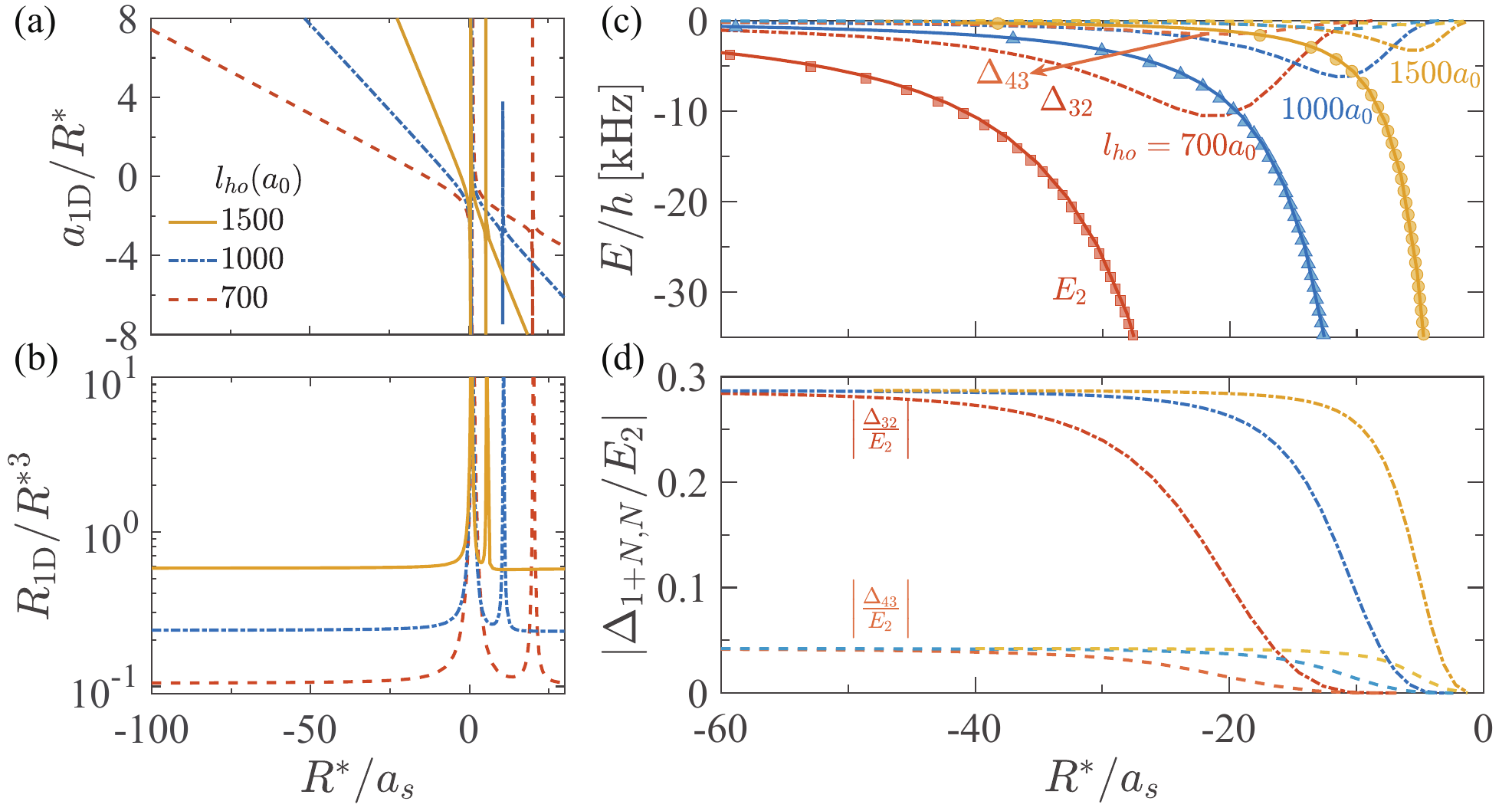}
	\caption{The same as Fig.\ref{fig_LiCr_q1d} except for $^6$Li-$^{40}$K mixture with 3D effective range $R^*=2400a_0$.  }
	\label{fig_LiK_q1d} 
\end{figure}

\end{widetext}

Next we proceed with q1D geometry. Figs.~\ref{fig_LiCr_q1d} and ~\ref{fig_LiK_q1d} show the q1D effective scattering and bound states for Li-Cr and Li-K systems, respectively. Again, multiple resonances appear in the 1D coupling constant $g_{\rm 1D} = -(\mu a_{\rm 1D})^{-1}$, originating from the multiple center-of-mass levels due to unequal trapping frequencies. Due to the large mass imbalance and large effective range of these systems, the resonance positions of $a_s$ (when scaled by $l_{ho}$) in Figs.~\ref{fig_LiCr_q1d} and ~\ref{fig_LiK_q1d} differ strongly from those in previous studies with much smaller mass imbalance and zero effective range\cite{NJP, split_CIR_expt}. 

Importantly, compared to the binding energies in q2D (Figs.~\ref{fig_LiCr_q2d}(c) and~\ref{fig_LiK_q2d}(c)), we find the q1D geometry can produce substantially deeper bindings for both $E_2$ and $\Delta_{1+N,N}$, as shown in Figs.~\ref{fig_LiCr_q1d}(c) and~\ref{fig_LiK_q1d}(c). For instance, at $l_{\text{ho}} = 700a_0$, the deepest $\Delta_{32}/h$ for  Li-Cr system is -0.92 kHz in q2D, but can reach -8.10 kHz in q1D; for  Li-K system, this value changes from -0.97 kHz in q2D to -10.61 kHz in q1D. Moreover, q1D yields a larger ratio $|\Delta_{32}/E_2|$ compared to the q2D case, as can be seen by comparing Fig.~\ref{fig_LiCr_q2d}(d) with Fig.~\ref{fig_LiCr_q1d}(d), and Fig.~\ref{fig_LiK_q2d}(d) with Fig.~\ref{fig_LiK_q1d}(d). Based on these observations, we can conclude that it is more favorable to detect $(1+2)$ trimer in q1D than in q2D.  For the $(1+3)$ tetramer, q1D enhances the relative binding energy $|\Delta_{43}|$, but produces a smaller $|\Delta_{43}/E_2|$ than q2D case. 

For the q1D Li-Cr system at $l_{\text{ho}} = 700a_0$, a favorable condition for detecting these clusters (see Figs.~\ref{fig_LiCr_q1d}(c,d)) lies within $R^*/a_s \in (-194.0, -127.1)$, where $\Delta_{32}/h \in (-1.79, -7.85)$kHz, $\Delta_{43}/h \in (-0.35, -1.13)$kHz, $|\Delta_{32}/E_2|\in (36\%, 20\%)$, and $|\Delta_{43}/E_2|\in(7\%,3\%)$. For the q1D Li-K system, the favorable condition (see Figs.~\ref{fig_LiK_q1d}(c,d)) is within $R^*/a_s \in (-53.3, -25.7)$, with $\Delta_{32}/h \in (-1.41, -8.81)$kHz, $\Delta_{43}/h \in (-0.21, -1.23)$kHz, and $|\Delta_{32}/E_2|\in(28\%, 20\%)$, and $|\Delta_{43}/E_2|\in (4\%, 3\%)$.


\section{Summary} \label{summary}

In summary, we have studied the effective scattering of two heteronuclear atoms in quasi-low dimensions under harmonic confinements with unequal frequencies for different atomic species. In this regime, the COM motion is strongly coupled to the relative motion, leading to multiple scattering resonances in the reduced low-dimensional couplings. By incorporating the finite effective range characteristic of ultracold heteronuclear mixtures, we have derived and computed the effective low-dimensional scattering parameters as functions of the 3D scattering parameters and confinement strength. Based on these reduced parameters, we have evaluated the relative binding energies of universal $(1+N)$ clusters in both q2D and q1D, using realistic Li-Cr and Li-K mixtures as concrete examples, thereby identifying the optimal conditions for their practical detection. Our results indicate that, for the same confinement length, it is more favorable to detect two-body bound state and $(1+2)$ trimer in q1D than in q2D.

This work paves the way for detecting universal few-body bound states in quasi-low-dimensional mass-imbalanced atomic mixtures with arbitrary harmonic confinements. The successful detection of these clusters will serve as a crucial first step toward exploring their associated many-body phases, which feature intriguing quantum correlations in few-body channels \cite{Parish3, Parish4, Naidon, mass_polaron, QSF}.

\bigskip

{\it Note added:} We note that a recent preprint independently solves the effective scattering in q2D with unequal trapping frequencies, using a different approach from ours~\cite{wang}. The obtained effective parameters are consistent with ours.

\bigskip

{\it Data availability.} The data that support the findings of this article are openly available~\cite{data}.

\bigskip

{\it Acknowledgements.} This work is supported by National Natural Science Foundation of China (12525412, 92476104, 12134015) and Quantum Science and Technology-National Science and Technology Major Project (2024ZD0300600). 
 
 
\appendix

\section{Details of quasi-2D scattering} \label{appendix_q2D}
In this section, we explicitly detail the calculation of the matrix element $\mathcal{M}_{NN'}$ in quasi-2D scattering and the regularization of the ultraviolet (UV) divergence arising from the 3D contact interaction. For convenience, we denote the three terms in Eq.~(\ref{M_element}) as $I_1$, $I_2$, and $I_3$, respectively.

As seen from Eq.~(\ref{range}), the 3D contact interaction diverges at large momenta. To regularize this UV divergence in the $1/g$ term, we utilize the integral identity
\begin{eqnarray} 
	\frac{1}{x} = -\int_0^{+\infty} e^{xt} dt, \quad x<0
	\label{invx}
\end{eqnarray} 
to integrate out the summation over the 3D momentum $\mathbf{Q}$, yielding
\begin{eqnarray} 
	-\frac{1}{V}\sum_{\cp Q} \frac{2\mu}{{\cp Q}^2} 
	=-\int_0^{+\infty} dt \left( \frac{\mu}{2\pi t} \right)^{\frac{3}{2}}  .
\end{eqnarray} 
The high-energy UV divergence manifests when the time variable $t$ approaches zero. To isolate this divergent contribution, we introduce a small short-time cutoff $t_0$. Consequently, the first term in Eq.~(\ref{M_element}) can be separated as
\begin{eqnarray} 
	I_1=D_1 + \left[\frac{\mu}{2\pi a_s(E_N)}
	-\int_{t_0}^{+\infty}dt \left( \frac{\mu}{2\pi t} \right)^{\frac{3}{2}} \right] \delta_{NN'},
\end{eqnarray} 
where the purely divergent part is captured by
\begin{eqnarray} 
	D_1 =
	-\int_0^{t_0}dt \left( \frac{\mu}{2\pi t} \right)^{\frac{3}{2}} \delta_{NN'}.
	\label{D1}
\end{eqnarray} 

For the third term $I_3$ in Eq.~(\ref{M_element}), assuming a large momentum cutoff $\tilde{\Lambda}$, we can safely approximate $\ln\tilde{\Lambda}^2 \approx \ln(\tilde{\Lambda}^2 + 1)$ to rewrite this term as a summation over the 2D momentum $\mathbf{k}$:
\begin{eqnarray} 
	\frac{\mu}{2\pi} \ln\tilde{\Lambda}^2
	\approx\frac{\mu}{2\pi} \ln(\tilde{\Lambda}^2 +1)
	=\frac{1}{S}\sum_{\bf k}
	\frac{2\mu}{k^2+1/l_0^2} .
\end{eqnarray} 
Resorting again to Eq.~(\ref{invx}) and integrating out $\mathbf{k}$, $I_3$ is expressed as
\begin{eqnarray} 
	I_3
	=&&\frac{\mu}{2\pi}\int_0^{+\infty}d t\,
	\frac{e^{-\frac{1}{2\mu l_0^2}t}}{t}F_{N; 0,0} F^*_{N'; 0,0}\notag\\
	=&&D_2 + \frac{\mu}{2\pi}\int_{t_0}^{+\infty}d t\,
	\frac{e^{-\frac{1}{2\mu l_0^2}t}}{t}F_{N; 0,0} F^*_{N'; 0,0},
	\label{open}
\end{eqnarray} 
where the short-time integral 
\begin{eqnarray} 
D_2 = \frac{\mu}{2\pi}\int_0^{t_0} d t\,
	\frac{e^{-\frac{1}{2\mu l_0^2}t}}{t}F_{N; 0,0} F^*_{N'; 0,0}
	\label{D2}
\end{eqnarray} 
contains the UV divergence as $t \to 0^+$. Here, $F_{N; 0,0}$ can be evaluated analytically as
\begin{eqnarray} 
	F_{N; 0,0}=\left(
	\frac{m_h\omega_h m_l\omega_l}{M\omega_M}
	\right)^{1/4}
	\frac{(-2)^{\frac{N}{2}} \Gamma(\frac{N+1}{2}) _2F_1(-\frac{N}{2},\frac{1}{2};\frac{1}{2};\frac{1}{2\alpha})}
	{\pi^{3/4}\sqrt{2\alpha N!}}\notag\\
\end{eqnarray} 
with Gauss hypergeometric function $_2F_1(a,b;c;z)$ and the dimensionless parameter
\begin{eqnarray} 
	\alpha=\frac{1}{4}\left(
	1 + \frac{m_h\omega_h + m_l\omega_l}{M\omega_M}
	\right).
\end{eqnarray} 

We now demonstrate that the UV divergences identified above are exactly canceled by the contributions from $I_2$. By applying Eq.~(\ref{invx}) and integrating out $\mathbf{k}$, $I_2$ separates into two parts, $I_2 = J_1 + J_2$, where $J_1$ is expressed in terms of the 1D imaginary-time propagator of a harmonic oscillator $g_{1D}^{\sigma}$. Explicitly,
\begin{widetext}
\begin{align} 
	J_1 = &\frac{\mu}{2\pi}
	\int_0^{+\infty}d t\, \frac{e^{Et}}{t} \bigg[\int d Z\int d Z'   
	\Phi_N(Z) \Phi_{N'}(Z')
	g_{1D}^h(Z,Z',t) g_{1D}^l(Z,Z',t)\bigg],\\
	J_2=&-\frac{\mu}{2\pi}\int_0^{+\infty} dt\,
	\frac{e^{\frac{q^2}{2\mu}t}}{t}
	F_{N;0,0}F^*_{N';0,0},\label{eq:J2}\\
	g_{1D}^{\sigma} 
	=&\sum_{n=0}^{+\infty}
	\phi_{n}^{\sigma}(Z)\phi_{n}^{\sigma}(Z') e^{-\epsilon_{n}^{\sigma} t}= \sqrt{B_{\sigma}(t)}
	\exp\left\{
	\frac{m_{\sigma}\omega_{\sigma}\left[
		2ZZ'-\cosh(\omega_{\sigma} t) (Z^2+{Z'}^2)
		\right]}{2\sinh(\omega_{\sigma} t)}
	\right\},
\end{align} 
\end{widetext}
where $B_{\sigma}(t) =m_{\sigma}\omega_{\sigma}\sinh^{-1}(\omega_{\sigma} t)/(2\pi)$.
Here, $\Phi_N$ is the eigenstate of $H_M$ satsfying $H_M\Phi_N = \epsilon_N\Phi_N$ with the eigenvalue $\epsilon_N=(N+1/2)\omega_M$, having the specific form 
\begin{eqnarray} 
	\Phi_N(Z) = \left( \frac{M\omega_M}{\pi}
	\right)^{\frac{1}{4}} 
	\frac{e^{-\frac{M \omega_MZ^2}{2}}}{\sqrt{2^NN!}}
	H_N\left( \sqrt{M\omega_M}Z \right).
\end{eqnarray} 
Similarly, $\phi_n^{\sigma}(z)$ is the single-particle eigenstate for the $\sigma$-atom corresponding to the eigenvalue $\epsilon^{\sigma}_{n}=(n+\frac{1}{2})\omega_{\sigma}$:
\begin{eqnarray} 
	\phi_n^{\sigma}(z) = \left( \frac{m_{\sigma}\omega_{\sigma}}{\pi} 
	\right)^{\frac{1}{4}}
	\frac{e^{-\frac{m_{\sigma}\omega_{\sigma} z^2}{2}}}{\sqrt{2^n n!}}
	 H_n(\sqrt{m_{\sigma}\omega_{\sigma}}z).
\end{eqnarray} 

The term $J_2$ diverges in both the short-time ($t\to 0^+$) and long-time ($t\to +\infty$) limits. We isolate these divergences by dividing the integration range using the short-time cutoff $t_0$ and a long-time cutoff $t_c$:
\begin{eqnarray} 
	J_2= \tilde{D}_2 + D_3
	-\frac{\mu}{2\pi}\int_{t_0}^{t_c} dt\,
	\frac{e^{\frac{q^2}{2\mu}t}}{t}
	F_{N;0,0}F^*_{N';0,0}.
\end{eqnarray} 
The short-time ($\tilde{D}_2$) and long-time ($D_3$) divergent contributions correspond to 
\begin{eqnarray} 
	\tilde{D}_2 =&&
	-\frac{\mu}{2\pi}\int_0^{t_0}  dt\,
	\frac{e^{\frac{q^2}{2\mu}t}}{t}
	F_{N;0,0}F^*_{N';0,0},\\
	D_3=&&-\frac{\mu}{2\pi}\int_{t_c}^{+\infty}  dt\,
	\frac{e^{\frac{q^2}{2\mu}t}}{t}
	F_{N;0,0}F^*_{N';0,0}.
\end{eqnarray} 
The short-time divergence $\tilde{D}_2$ can be explicitly canceled by incorporating $D_2$ from Eq.~(\ref{open}). Expanding the exponential for small $t_0$ leads to
\begin{eqnarray} 
	D_2+\tilde{D}_2\approx -\frac{\mu}{2\pi}
	\left(\frac{q^2}{2\mu}+\frac{1}{2\mu l_0^2} \right) t_0
	F_{N;0,0}F^*_{N';0,0},
\end{eqnarray} 
which is finite and well-defined.

Similarly, the integral in $J_1$ diverges at small $t$. We denote the short-time contribution (from $0$ to $t_0$) as $\tilde{D}_1$:
\begin{eqnarray} 
	\tilde{D}_1=&&\frac{\mu}{2\pi}
	\int_0^{t_0}d t\, \frac{e^{Et}}{t} \bigg[\int d Z\int d Z'  
	\Phi_N(Z) \Phi_{N'}(Z')\notag\\
	&&
	g_{1D}^h(Z,Z',t) g_{1D}^l(Z,Z',t) \bigg],
\end{eqnarray} 
which will be removed by combining it with $D_1$. To proceed, we introduce a non-interacting Hamiltonian governing the  motion:
\begin{eqnarray} 
	H_0=K({\bf P},{\bf p})+V(Z,{\bf r})
\end{eqnarray} 
with the kinetic energy 
\begin{eqnarray} 
	K({\bf P},{\bf p})=\frac{{\bf P}^2}{2M} + \frac{{\bf p}^2}{2\mu}
\end{eqnarray} 
and the potential energy 
\begin{eqnarray} 
	V(Z,{\bf r})=V_h(Z+\frac{m_l}{m_h+m_l}z) +V_l(Z-\frac{m_h}{m_h+m_l}z).\notag\\
\end{eqnarray} 
Note that ${\bf P}$ and ${\bf p}$ are the momentum operators for COM motion in the longitudinal direction and relative motion in 3D space, respectively, $Z$ and ${\bf r}$ are the COM coordinate along $z$-direction and 3D relative coordinate. After some algebra, $\tilde{D}_1$ can be compactly rewritten as
\begin{eqnarray} 
	\tilde{D}_1=\int_0^{t_0}d t\, \int d Z d Z' e^{Et} 
	\bra{N}\ket{Z} G_t(Z,Z') \bra{Z'}\ket{N'}\notag\\
	\label{D1_tl}
\end{eqnarray} 
with the Green's function evaluated at $\bm{r}=\mathbf{0}$:
\begin{eqnarray} 
	G_t(Z,Z') = \bra{Z}e^{-H_0t}\ket{Z'},
\end{eqnarray} 
where we take $\ket{Z}=\ket{Z,\bm{r}={\bf 0}}$.

To evaluate $G_t(Z,Z')$ for small $t$, we treat the kinetic part $\mathcal{K}(Z,Z')=\bra{Z} e^{-Kt} \ket{Z'}$ and the potential part $\mathcal{V}(Z,Z')=\bra{Z} e^{-Vt} \ket{Z'}$ via a first-order Trotter expansion.
Evaluating the standard Gaussian integral over momenta, the kinetic part yields
\begin{eqnarray} 
	\mathcal{K}(Z,Z')
	=&&\sqrt{\frac{M}{2\pi t}}\left(\frac{\mu}{2\pi t}\right)^{\frac{3}{2}}
	e^{-\frac{M(Z-Z')^2}{2t}} \notag\\
	=&&\left(\frac{\mu}{2\pi t}\right)^{\frac{3}{2}}
	\sum_{P}	e^{iP(Z-Z')-\frac{P^2}{2M}t}\notag\\
	=&&\left(\frac{\mu}{2\pi t}\right)^{\frac{3}{2}}
	\left[
	\delta(Z-Z')
	- \bra{Z}\frac{{\bf P}^2}{2M}t\ket{Z'} \right],\notag\\
\end{eqnarray} 
where we Taylor-expanded with respect to $t$. Utilizing the approximation $G_t(Z,Z') \approx(1-Vt)\bra{Z}e^{-Kt}\ket{Z'}$ and substituting  back into Eq.~(\ref{D1_tl}) and evaluating the action of $H_M$ given by Eq.~(\ref{HM_2d}), we finally obtain $\tilde{D}_1$. Combining this with $D_1$, the leading divergence precisely vanishes, leaving only a finite terms. Specifically,
\begin{widetext}
\begin{align} 
	&G_t(Z,Z') 
	\approx [ 1-V(Z,{\bf 0})t ] \mathcal{K}(Z,Z')\notag\\
	&\qquad\qquad \approx \left(\frac{\mu}{2\pi t}\right)^{\frac{3}{2}} \delta(Z-Z')
	-\left(\frac{\mu}{2\pi }\right)^{\frac{3}{2}}  \frac{1}{\sqrt{t}} \left[
	V(Z,{\bf 0})\delta(Z-Z') + \bra{Z}\frac{{\bf P}^2}{2M}\ket{Z'}
	\right]\notag\\
	&\qquad\qquad=\left(\frac{\mu}{2\pi t}\right)^{\frac{3}{2}} 
	\delta(Z-Z')
	-   \left(\frac{\mu}{2\pi }\right)^{\frac{3}{2}}  \frac{1}{\sqrt{t}}
	\bra{Z} H_M \ket{Z'} ,
	\label{Gt}\\
	&\tilde{D}_1
	=\left(\frac{\mu}{2\pi t }\right)^{\frac{3}{2}}\int_0^{t_0}d t\, e^{Et} 
	(1-t\epsilon_{N})\delta_{NN'}
	\approx\int_0^{t_0}d t\,  
	\left(\frac{\mu}{2\pi t}\right)^{\frac{3}{2}} \delta_{NN'}
	+ 2\left(\frac{\mu}{2\pi}\right)^{\frac{3}{2}} \left[
	(E-\epsilon_{N}) \sqrt{t_0}
	-\frac{E \epsilon_{N}}{3} t_0^{\frac{3}{2}}	
	\right]\delta_{NN'},\\
&	D_1 + \tilde{D}_1
	= 2\left(\frac{\mu}{2\pi}\right)^{\frac{3}{2}} \left[
	(E-\epsilon_{N}) \sqrt{t_0}
	-\frac{E \epsilon_{N}}{3} t_0^{\frac{3}{2}}	
	\right]\delta_{NN'}.
\end{align} 
\end{widetext}
Furthermore, the remaining integral in $J_1$ from $t_0$ to $+\infty$, denoted as $C_1$ with the spatial integration $\mathcal{F}_{NN'}(t)$, which can be evaluated as
\begin{widetext}
\begin{align} 
	&C_1=\frac{\mu}{2\pi} \int_{t_0}^{+\infty}d t\, 	\frac{e^{Et}}{t} \left[
	\int d Z\int d Z' \Phi_N(Z) \Phi_{N'}(Z')
	g_{1D}^h(Z,Z',t) g_{1D}^l(Z,Z',t)\right]
	=\frac{\mu}{2\pi}\int_{t_0}^{+\infty}d t\,
	\mathcal{F}_{NN'}(t) h(t),\\
	&\mathcal{F}_{NN'}(t) 
	=\int_{-\infty}^{+\infty} d Z  d Z'\Phi_N(Z)\Phi_{N'}(Z') 
	\exp\left[
	g(t) ZZ'
	-f(t) \frac{Z^2+{Z'}^2}{2}
	\right]\notag\\
	=&
	\begin{cases}
		\sqrt{\frac{[1-A^2(t)]^{N}}{2^{N+N'} N! N'!}}
		A(t)
		\int_{-\infty}^{+\infty} d Z
		\exp\left[
		-Z^2\left(
		\frac{M\omega_M}{A^2(t)} - \frac{g^2(t)A^2(t)}{4M\omega_M}
		\right)
		\right]
		H_N\left(
		\frac{g(t)A^2(t)}{2\sqrt{M\omega_M [1-A^2(t)]}} Z
		\right)
		H_{N'}(\sqrt{M\omega_M}Z),
		& A(t)\neq 1, \\
		(2^{N+N'} N! N'!)^{-\frac{1}{2}}\left[\frac{g(t)}{\sqrt{M\omega_M}}\right]^{N}
		\int_{-\infty}^{+\infty} d Z
		\exp\left[
		-Z^2\left(
		M\omega_M - \frac{g^2(t)}{4M\omega_M}
		\right)
		\right]
		H_{N'}(\sqrt{M\omega_M}Z) Z^N,
		& A(t)=1,
	\end{cases}
	\notag\\
	\label{q2deq:Ft1}
\end{align} 
\end{widetext}
with the auxiliary functions defined as
\begin{eqnarray} 
	g(t)=&&2\pi\left[B_h(t) + B_l(t)\right],\\
	f(t)=&&\frac{m_h\omega_h}{\tanh(\omega_h t)} + \frac{m_l\omega_l}{\tanh(\omega_l t)},\\
	A(t)=&&\sqrt{\frac{2M\omega_M}{f(t)+M\omega_M}},\\
	h(t)=&&\frac{e^{Et}}{2\pi t}
	\sqrt{\frac{m_h\omega_h m_l\omega_l}{\sinh(\omega_h t)\sinh(\omega_l t)}}.
\end{eqnarray} 
In the long-time limit ($t\to +\infty$), the imaginary-time propagator simplifies to
\begin{eqnarray} 
	g_{1D}^{\sigma} (t\to +\infty)
	\to \sqrt{\frac{m_{\sigma}\omega_{\sigma}}{\pi e^{\omega_{\sigma} t}}}
	\exp{\left[
	-\frac{m_{\sigma}\omega_{\sigma}(Z^2+{Z'}^2)}{2}
	\right]},\notag\\
\end{eqnarray} 
such that the long-time divergent part of $C_1$ (integrated from $t_c$ to $+\infty$) exactly cancels the divergence $D_3$ defined earlier.

Finally, combining all the finite contributions, we obtain the divergence-free matrix element $\mathcal{M}_{NN'}$:
\begin{widetext}
\begin{eqnarray} 
	\mathcal{M}_{NN'}=&& \left\{\frac{\mu}{2\pi a_s(E_N)}
	+2\left(\frac{\mu}{2\pi}\right)^{\frac{3}{2}} \left[
	(E-\epsilon_{N}) \sqrt{t_0}
	-\frac{E \epsilon_{N}}{3} t_0^{\frac{3}{2}}	
	\right]
	-\int_{t_0}^{+\infty}dt \left( \frac{\mu}{2\pi t} \right)^{\frac{3}{2}} \right\} \delta_{NN'} 
	-\frac{\mu}{2\pi}
	\left(\frac{q^2}{2\mu}+\frac{1}{2\mu l_0^2} \right) t_0
	F_{N;0,0}F^*_{N';0,0} \notag\\
	&&
	+ \frac{\mu}{2\pi}\int_{t_0}^{+\infty}d t\,
	\frac{e^{-\frac{1}{2\mu l_0^2}t}}{t}F_{N; 0,0} F^*_{N'; 0,0}
	-\frac{\mu}{2\pi}\int_{t_0}^{t_c} dt\,
	\frac{e^{\frac{q^2}{2\mu}t}}{t}
	F_{N;0,0}F^*_{N';0,0}
	+ \frac{\mu}{2\pi}\int_{t_0}^{t_c}d t\,
	\mathcal{F}_{NN'}(t) h(t).
\end{eqnarray} 
\end{widetext}

\section{Details of quasi-1D scattering} \label{appendix_q1D}
This appendix outlines the evaluation of the quasi-1D matrix element $\mathcal{M}_{\mathbf{NN'}}$. While the regularization scheme shares conceptual similarities with the quasi-2D geometry, the dimensional reduction significantly modifies the specific analytical forms of the divergences. The target matrix element is given by
\begin{equation}
	\mathcal{M}_{\bf NN'}=
	\frac{1}{g}\delta_{\bf NN'} - \sum_{{{\cp n}_h,{\cp n}_l}\neq{{\cp 0,\cp 0}}; k} \frac{ F_{\cp N; \cp n_h, \cp n_l}  F^*_{\cp N'; \cp n_h, \cp n_l}}{E-E_{{\cp n}_h, {\cp n}_l; k}}.
	\label{M_ele_q1d}
\end{equation}
Here the function $F_{\cp N; \cp n_h, \cp n_l}$  is defined as 
\begin{eqnarray} 
	F_{{\bf N};{\bf n}_h,{\bf n}_l} 
	= \bra{\bf N}\ket{{\bf n}_h,{\bf n}_l;k}
	=\int d\pmb{\rho} \Phi^*_{\bf N}(\pmb{\rho}) \phi^h_{{\bf n}_h}(\pmb{\rho})  \phi^l_{{\bf n}_l}(\pmb{\rho}),\notag\\
\end{eqnarray} 
where the spatial integration over the transverse plane is performed as  $\int d \pmb{\rho} = \int_0^{+\infty} \rho d \rho \int_0^{2\pi} d\phi$.

To systematically address the singularities, we label the two terms  on the right-hand side of Eq.~(\ref{M_ele_q1d}) by $I_1$ and $I_2$. The first term, $I_1$, behaves identically to its quasi-2D counterpart discussed in Appendix~\ref{appendix_q2D}, provided the substitution $\delta_{NN'} \to \delta_{\mathbf{NN'}}$. The second term separates into two parts, $I_2 = J_1+J_2$, given by
\begin{eqnarray} 
	J_1=&&\sqrt{\frac{\mu}{2\pi}}  \int_0^{+\infty}d t\,\frac{e^{Et} }{\sqrt{t}} \bigg[
	\int d \pmb{\rho} \int d \pmb{\rho'} 
	\Phi^*_{\bf N}(\pmb{\rho}) \Phi_{\bf N'}(\pmb{\rho'})\notag\\
	&&g^h_{\rm 2D} (\pmb{\rho},\pmb{\rho'},t)
	g^l_{\rm 2D} (\pmb{\rho},\pmb{\rho'},t) \bigg],\\
	J_2=&&-\sqrt{\frac{\mu}{2\pi}} \int_0^{+\infty}d t\,
	\frac{e^{\frac{q^2}{2\mu}t}}{\sqrt{t}}
	F_{\bf N;0,0}F^*_{\bf N';0,0}.
\end{eqnarray} 
These are expressed in terms of the 2D imaginary-time propagator of a harmonic oscillator in polar coordinates (with $\rho$ and $\phi$ being the radial distance and polar angle for $\boldsymbol{\rho}$):
\begin{eqnarray} 
	g_{\rm 2D}^{\sigma}
	=\sum_{\bf n}
	\phi_{\bf n}^{\sigma}(\pmb{\rho})\phi_{\bf n}^{{\sigma}*}(\pmb{\rho'}) e^{-\epsilon_{\bf n}^{\sigma} t}=g_1^{\sigma}(\rho,\rho',\delta\phi,t) \, g_2^{\sigma}(\rho,\rho',t),\notag\\
\end{eqnarray}
where $\delta\phi=\phi-\phi'$, and the auxiliary functions are defined as
\begin{eqnarray}
	&&g_1^{\sigma}(\rho,\rho',\delta\phi,t)
	= \sqrt{B_{\sigma} (t)}\,
	\exp\left[
	\frac{m_{\sigma} \omega_{\sigma} }{\tanh(\omega_{\sigma} t)}
	\rho \rho' \cos(\delta\phi)
	\right] \notag\\
\end{eqnarray}
and 
\begin{eqnarray}
	g_2^{\sigma}(\rho,\rho',t)
	= \sqrt{B_{\sigma}(t)}\,
	\exp\left[
	- \frac{m_{\sigma}\omega_{\sigma}}{2\tanh(\omega_{\sigma} t)} (\rho^2 + {\rho'}^2)
	\right].\notag\\
\end{eqnarray}
Here, $\Phi_{\bf N}$ is the eigenstate of the molecule Hamitionian $H_M$, satisfying $H_M \Phi_{\bf N} = \epsilon_{\bf N} \Phi_{\bf N}$ with the eigenvalue $\epsilon_{\bf N}= (2 N_r + \abs{M_a} +1)\omega_M$. It takes the specific form
\begin{eqnarray} 
	\Phi_{\bf N}(\pmb{\rho})
	= C_{\bf N} U_{\bf N} (\rho) V_{M_a}(\phi)
\end{eqnarray} 
with the radial wavefunction 
\begin{eqnarray} 
	U_{\bf N}(\rho) 
	= \rho^{\abs{M_a}} \, e^{-\frac{M\omega_M{\rho}^2}{2} } \,
	L_{N_r}^{\abs{M_a}} (M\omega_M {\rho}^2),
\end{eqnarray} 
the angular wavefucntion 
\begin{eqnarray} 
	V_{M_a}(\phi) = e^{i M_a \phi},
\end{eqnarray} 
and the normalization coefficient
\begin{eqnarray} 
	C_{\bf N} = \sqrt{\frac{N_r !}{\pi(N_r + \abs{M_a})!}} \,
	(M\omega_M)^{\frac{\abs{M_a}+1}{2}}.
\end{eqnarray} 
We have introduced the combined notation $\mathbf{N}=(N_r,M_a)$ denoting the radial and angular momentum quantum numbers, and $L_n^m(x)$ corresponds to the associated Laguerre polynomials.

Similarly, $\phi_{\bf n}^{\sigma}(\pmb{\rho})$
is the wavefunction of a 2D harmonic oscillator for the $\sigma$-atom, corresponding to the eigenvalue $\epsilon^{\sigma}_{\bf n} =(2n+\abs{m}+1)\omega_{\sigma}$:
\begin{eqnarray} 
	\phi_{\bf n}^{\sigma}(\pmb{\rho})
	= c^{\sigma}_{\bf n} u^{\sigma}_{\bf n}(\rho) v^{\sigma}_m(\varphi)
\end{eqnarray} 
with the radial part
\begin{eqnarray} 
	u^{\sigma}_{\bf n}(\rho) 
	= \rho^{\abs{m}} e^{-\frac{m_{\sigma}\omega_{\sigma}\rho^2}{2}} L_n^{\abs{m}}(m_{\sigma}\omega_{\sigma}\rho^2),
\end{eqnarray} 
the angular part
\begin{eqnarray} 
	v^{\sigma}_{m}(\varphi) 
	= e^{im\varphi},
\end{eqnarray} 
and the normalization coefficient
\begin{eqnarray} 
	c^{\sigma}_{\bf n} = \sqrt{\frac{n!}{\pi (n+\abs{m})!}}
	(m_{\sigma}\omega_{\sigma})^{\frac{\abs{m}+1}{2}},
\end{eqnarray} 
where $\mathbf{n}=(n,m)$ represents the corresponding quantum numbers.

A key distinction from the quasi-2D geometry emerges in the short-time behavior. For the quasi-1D configuration, the integral $J_2$ converges as $t \to 0^+$, indicating the absence of an ultraviolet (UV) divergence in this specific term. However, it still diverges in the long-time limit ($t \to +\infty$). Furthermore, since $F_{\mathbf{N;0,0}}$ can be evaluated as
\begin{eqnarray} 
	F_{\bf N;0,0}
	=f_{N_r,0;{\bf  0,0}}	\delta_{M_a,0},
\end{eqnarray} 
where 
\begin{eqnarray} 
	f_{N_r,0;{\bf 0,0}}
	=2\sqrt{ \frac{m_l\omega_l m_h\omega_h  M\omega_M}{\pi} }
	\frac{(m_h\omega_h+m_l\omega_l - M\omega_M)^{N_r}}{(m_h\omega_h+m_l\omega_l + M\omega_M)^{N_r+1}}\notag\\
\end{eqnarray} 
with the aid of the integral identity
\begin{eqnarray} 
	\int_0^{+\infty} e^{-ax} L_n(x) d x=\frac{(a-1)^n}{a^{n+1}},
\end{eqnarray} 
we can explicitly calculate $J_2$ as
\begin{eqnarray} 
	J_2 = -\sqrt{\frac{\mu}{2\pi}} \int_0^{+\infty}d t\,
	\frac{e^{\frac{q^2}{2\mu}t}}{\sqrt{t}}\,
	f_{N_r,0;{\bf 0,0}} \, f_{N'_r,0;{\bf 0,0}} \,
	\delta_{M_a,0}  \,
	\delta_{M'_a,0}.\notag\\
\end{eqnarray} 

The short-time divergence also arises from $J_1$. We denote this short-time contribution (integrated from $0$ to $t_0$) as $\tilde{D}_1$, which will be removed by incorporating the divergent term $D_1$ from $I_1$ [analogous to Eq.~(\ref{D1})]. Following the procedure in Appendix~\ref{appendix_q2D}, we introduce a non-interacting Hamiltonian governing the two-body motion:
\begin{eqnarray} 
	H_0 = K({\bf P}, {\bf p}) + V(\pmb{\rho}, {\bf r})
\end{eqnarray} 
comprising the kinetic term 
\begin{eqnarray} 
	K({\bf P}, {\bf p}) 
	=\frac{{\bf P}^2}{2M} + \frac{{\bf p}^2}{2\mu},
\end{eqnarray} 
and the potential term
\begin{eqnarray} 
	V(\pmb{\rho}, {\bf r})
	=&&V_h( X+ \frac{\mu}{m_h} x,  Y+ \frac{\mu}{m_h} y )\notag\\
	&&+ V_l( X- \frac{\mu}{m_l} x,  Y- \frac{\mu}{m_l} y ).
\end{eqnarray} 
Note that ${\bf P}$ and ${\bf p}$ here are the momentum operators for COM motion in the transverse plane and the relative motion in 3D space, respectively. Correspondingly,  $\pmb{\rho}$ is the COM coordinate in the $xy$-plane and ${\bf r}$ is the 3D relative coordinate. After some algebra, $\tilde{D}_1$ can be compactly rewritten as
\begin{eqnarray} 
	\tilde{D}_1=\int_0^{t_0}d t\, \int d \pmb{\rho}  d \pmb{\rho'} e^{Et} 
	\bra{\bf N}\ket{\pmb{\rho}}
	G_t(\pmb{\rho}, \pmb{\rho'})
	\bra{\pmb{\rho'}}\ket{\bf N'},
	\label{D1_tl_1d}
\end{eqnarray} 
with the Green's function at ${\bf r}={\bf 0}$:
\begin{eqnarray} 
	G_t(\pmb{\rho}, \pmb{\rho'}) 
	= \bra{\pmb{\rho}}e^{-H_0t}\ket{\pmb{\rho'}}.
\end{eqnarray} 
By expanding this Green's function asymptotically for small $t$, we find
\begin{widetext}
\begin{eqnarray} 
	G_t(\pmb{\rho}, \pmb{\rho'}) 
	= \left(\frac{\mu}{2\pi t}\right)^{\frac{3}{2}} \delta(\pmb{\rho}-\pmb{\rho'})
	-\left(\frac{\mu}{2\pi}\right)^{\frac{3}{2}} \frac{1}{\sqrt{t}}
	\bra{\pmb{\rho}}H_M\ket{\pmb{\rho'}},
	\label{Gt_1d}
\end{eqnarray} 
\end{widetext}
where $H_M$ is the molecule Hamiltonian for the  quasi-1D configuration  given by Eq.~(\ref{HM_1d}).
Plugging Eq.~(\ref{Gt_1d}) into Eq.~(\ref{D1_tl_1d}) and acting $H_M$ on the eigenstates leads to
\begin{eqnarray} 
	\tilde{D}_1=&&\int_0^{t_0}\dd t\,  
	\left(\frac{\mu}{2\pi t}\right)^{3/2} \delta_{\bf NN'}\notag\\
	+&& 2 \left(\frac{\mu}{2\pi}\right)^{\frac{3}{2}} \Bigg[
	(E-\epsilon_{\bf N}) \sqrt{t_0}
	-\frac{E \epsilon_{\bf N}}{3} t_0^{\frac{3}{2}}	
	\Bigg]\delta_{\bf NN'}.\notag\\
\end{eqnarray} 
Upon adding the intrinsic divergence $D_1$ from $I_1$, the singularity in $\tilde{D}_1$ is perfectly canceled, producing a finite residual identical in form to the quasi-2D result:
\begin{eqnarray} 
D_1 +	\tilde{D}_1= 2 \left(\frac{\mu}{2\pi}\right)^{\frac{3}{2}} \Bigg[
	(E-\epsilon_{\bf N}) \sqrt{t_0}
	-\frac{E \epsilon_{\bf N}}{3} t_0^{\frac{3}{2}}	
	\Bigg]\delta_{\bf NN'}.\notag\\
\end{eqnarray} 

The remaining integral in $J_1$ from $t_0$ to $+\infty$, denoted as $C_1$, with the spatial integration $\mathcal{F}_{N_r,\abs{M_a}; N'_r, \abs{M_a}}(t)$. They  can be explicitly evaluated as
\begin{widetext}
\begin{align} 
	&C_1=\sqrt{\frac{\mu}{2\pi}}  \int_{t_0}^{+\infty}d t\,\frac{e^{Et} }{\sqrt{t}} \bigg[
	\int d \pmb{\rho}  d \pmb{\rho'} 
	\Phi^*_{\bf N}(\pmb{\rho}) \Phi_{\bf N'}(\pmb{\rho'})g^h_{\rm 2D} (\pmb{\rho},\pmb{\rho'},t)
	g^l_{\rm 2D} (\pmb{\rho},\pmb{\rho'},t) \bigg]
	=\sqrt{\frac{\mu}{2\pi}}\int_{t_0}^{+\infty} \dd t\,
	\mathcal{F}_{N_r,\abs{M_a}; N'_r, \abs{M_a}}(t) \, \delta_{M_a,M'_a},\notag\\
	&\mathcal{F}_{N_r,\abs{M_a}; N'_r, \abs{M_a}}(t)=
	\begin{cases}
		\displaystyle 2\pi (M\omega_M)^{\abs{M_a}} 	
		\sqrt{\frac{N_r! N'_r!}{(N_r+\abs{M_a})! (N'_r+\abs{M_a})!}}
		\frac{e^{Et}}{\sqrt{t}} B_h B_l\,
		(1-A^2)^{N'_r} A^2\,
		h^{\abs{M_a}} \\
		\displaystyle \quad \times \int d \rho\, {\rho}^{2\abs{M_a}+1}
		e^{- \frac{M\omega_M}{A^2}\left( 1 - h^2 \right) {\rho}^2}
		L_{N_r}^{\abs{M_a}} (M\omega_M{\rho}^2)
		L_{N'_r}^{\abs{M_a}}\left(
		\frac{M\omega_M h^2}{1-A^2} {\rho}^2
		\right), & A(t)\neq 1,\\[15pt]
		\displaystyle 2\pi (-1)^{\abs{M_a}}(-M\omega_M)^{N'_r+\abs{M_a}}
		\sqrt{\frac{N_r!}{(N_r+\abs{M_a})! (N'_r+\abs{M_a})!}}
		\frac{e^{Et}}{\sqrt{t}} B_h B_l\,
		\tilde{h}^{2N'_r+\abs{M_a}}\\
		\displaystyle \quad \times \int d \rho\, {\rho}^{2N'_r+2\abs{M_a}+1}
		e^{-M\omega_M\left( 1 - \tilde{h}^2 \right) {\rho}^2}
		L_{N_r}^{\abs{M_a}}(M\omega_M {\rho}^2), & A(t)=1.
	\end{cases}
\end{align}
\end{widetext} 
Here, the auxiliary functions are defined as
\begin{eqnarray} 
	h(t) =&& \frac{gA^2}{2M\omega_M},\\
	\tilde{h}(t) =&& \frac{g}{2M\omega_M}.
\end{eqnarray} 
As previously stated, the long-time divergence present in $C_1$ (for $t \to +\infty$) precisely cancels the long-time divergence in $J_2$.

Gathering all the regularized pieces together, the physical, divergence-free matrix element is ultimately cast into the form:
\begin{widetext}
	\begin{eqnarray} 
		\mathcal{M}_{\bf NN'}=&& \left\{\frac{\mu}{2\pi a_s(E_{\cp N})}
		+2\left(\frac{\mu}{2\pi}\right)^{\frac{3}{2}} \left[
		(E-\epsilon_{\cp N}) \sqrt{t_0}
		-\frac{E \epsilon_{\cp N}}{3} t_0^{\frac{3}{2}}	
		\right]
		-\int_{t_0}^{+\infty}dt \left( \frac{\mu}{2\pi t} \right)^{\frac{3}{2}} \right\} \delta_{\bf NN'} \notag\\
		&&
		-\sqrt{\frac{\mu}{2\pi}} \int_0^{t_c}d t\,
		\frac{e^{\frac{q^2}{2\mu}t}}{\sqrt{t}}\,
		f_{N_r,0;{\bf 0,0}} \, f_{N'_r,0;{\bf 0,0}} \,
		\delta_{M_a,0}  \,
		\delta_{M'_a,0}
		+ \sqrt{\frac{\mu}{2\pi}}\int_{t_0}^{t_c} \dd t\,
		\mathcal{F}_{N_r,\abs{M_a}; N'_r, \abs{M_a}}(t) \, \delta_{M_a,M'_a}.
	\end{eqnarray} 
\end{widetext}
Here, the delta functions result from the conservation of the total angular momentum along the confinement axis due to the rotational symmetry of the transverse confinement; hence, for the $s$-wave incident channel in the transverse ground state, only states with $M_a=0$ contribute.

\end{document}